\documentstyle[aps,pre,epsf]{revtex}

\begin{document}
\title{
\draft
Stochastic Resonance in a Dipole
}
\author{
J. M. G. Vilar, A. P\'erez-Madrid, and J. M. Rub\'{\i}
}
\address{
Departament de F\'{\i}sica Fonamental, Facultat de
F\'{\i}sica, Universitat de Barcelona, Diagonal 647,
E-08028 Barcelona, Spain 
\date{\today}
}
\maketitle
\widetext
\begin{abstract}
\leftskip 54.8pt
\rightskip 54.8pt
We show that the dipole, a system usually proposed to model relaxation
phenomena,  exhibits  a  maximum  in the  signal-to-noise  ratio  at a
non-zero  noise level, thus  indicating  the  appearance of stochastic
resonance.  The  phenomenon
occurs in two  different  situations,  i.e.
when the minimum of the potential of the dipole  remains fixed in time
and when it switches  periodically between two equilibrium points.
We have also found that the  signal-to-noise
ratio has a maximum for a certain  value of the  amplitude of the
oscillating field. 
\end{abstract}

\pacs{PACS numbers: 05.40.+j, 41.90.+e, 82.70.Dd}

The phenomenon of stochastic  resonance (SR)
\cite{Benzi,Maki1,Maki2,libi,JSP,Moss,Wies,Wiese,thre,phi4}
may be  characterized by
the  appearance  of a maximum  in the  output  signal-to-noise  ratio
(${\rm SNR}$) at an optimized  non-zero  noise level.  Although early
studies of SR were restricted to bistable systems, later developments
have shown that this  phenomenon  is also present in a wider class of
situations,  including  threshold devices and fire-and-reset  models.
In all these cases, a  combination  of the periodic  input signal and
the  presence  of an  optimized  noise  may give  rise to SR when the
system crosses a barrier or a threshold in a coherent  fashion.  This
phenomenon  has  been  found in many  particular  realizations  among
others a Schmitt  trigger \cite{tri},  lasers\cite{Maki1},
neurons \cite{neu1,neu2}  and  magnetic
particles \cite{Agus}.

In  spite  of the  efforts  devoted  to  show  the  ubiquity  of  the
phenomenon in many branches of physics,  chemistry and biology, there
is an aspect which has not been  considered  up to now.  It refers to
the possibility
of the appearance of a maximum in the SNR \cite{bbbt} during
the relaxation
process of a system towards a single minimum, due to the combined
action  of the  noise and the  periodic  signal. 
In this  paper  we
address this possibility  precisely,  which opens new applications of
SR in the domain of the  relaxation  phenomena.  To be more specific,
consider a system in an equilibrium  state, stable or unstable, where
no motion is  observed.  Addition  of noise  removes  the system from
this  situation in such a way that a force field acts on it.  If this
force varies  periodically in time, then SR may occur.  As a physical
realization  of the  phenomenon we have shown that SR may take
place in a single dipole under a periodic external field.

Let us consider the  relaxational  dynamics of a single  dipole
\cite{Deb,Deb2} in an
external  field  which  may be  described  by the  Langevin  equation
governing  the  dynamics  of the  unit  vector  $\hat  R$  along  the
direction of the dipole moment
\begin{equation} \label{eq:a1}
{d \hat R \over dt} =
\left(\vec h(t)
\times \hat R + \vec \xi(t) \right) \times \hat R \;\; .
\end{equation}
Here the total field  $\vec  h(t)=k(1+\alpha\sin(\omega_0t))\hat  z$,
with  $\hat  z$  a  unit  vector,  consists  of a  constant  plus  an
oscillating  field and it is characterized  by the parameters $k$ and
$\alpha$.  The  noise  term  $\vec  \xi(t)$  may  originate  from the
presence of an external  random  field or from thermal  fluctuations.
It is assumed to be  Gaussian  white  noise with zero mean and second
moment $\left<\xi_i(t)\xi_j(t+\tau)\right>=D\delta_{ij}\delta(\tau)$,
defining the noise level $D$.

The  potential  energy of the  dipole is  $V(\theta)=-h(t)cos\theta$,
where  $\theta$ is the angle between its direction  and the direction
of the  external  field  ($\cos(\theta)=\hat  z \cdot  \hat R$).  Two
qualitatively  different situations may occur, one for small ($\alpha
< 1$) and another for high ($\alpha > 1$) oscillating fields.  In the
first case, the  potential has a minimum which remains  fixed in time
at $\theta=0$ (Fig.  \ref{fig1}a).  In the absence of noise, once the
dipole  relaxes  toward this minimum no motion is  observed,  whereas
when  noise is added,  the  motion is  modulated  by the  oscillating
field.  In the second case (Fig.  \ref{fig1}b),  the  position of the
minimum switches  periodically in time between the equilibrium points
$\theta=0$ and  $\theta=\pi$.  In the absence of noise, starting from
any  arbitrary  initial  condition  ($\theta_0  \neq \pi$) the dipole
relaxes towards $\theta=0$, in spite of this point becoming stable or
unstable  periodically.  When a small  amount  of  noise  acts on the
dipole, it may leave the unstable equilibrium point toward the stable
one.  We could easily infer that the output signal has a maximum as a
function  of $D$ but it is not a  trivial  matter  to  elucidate  the
behavior of the SNR,  because both signal and noise go to zero as $D$
decreases.

Since we are concerned with the magnitude $\cos(\theta)$, which for a
magnetic  dipole  corresponds  to the component of the  magnetization
along the field, the averaged power spectrum is given by
\begin{equation} \label{eq:a2}
P(\omega) \equiv  \int_{-\infty}^{\infty}ds e^{i\omega s}
\int_0^{2\pi/\omega_0} 
\left<\cos(\theta_{t+s})\cos(\theta_{t})\right>dt \;\; .
\end{equation}
Analitycal  studies of the system, using linear response theory (LRT)
which is valid for small  fields, show a maximum in the signal,  i.e.
in  the  susceptibility,  as a  function  of  the  noise  level  $D$.
However, the ${\rm SNR}$ is a  monotonically  decreasing  function of
$D$.  One  finds
\begin{equation}   \label{eq:a3}
{\rm  SNR}  = L(\mu)k\alpha^2    \;\;   ,
\end{equation}
obtained from the fluctuation-dissipation  theorem \cite{Dau}
and the expression
for the complex susceptibility \cite{Cof}
\begin{equation} \label{eq:a4}
\chi(\omega)={{\mu \over  k}   L^\prime(\mu)   \over  1  +
i\omega\tau(\mu)}   \;\;  ,
\end{equation}
where $\tau(\mu)={\mu^2  L^\prime(\mu) / 2kL(\mu)}$ is the relaxation
time and  $L(\mu)=\coth(\mu)-1/\mu$  is the Langevin  function,  with
$\mu\equiv 2h(t)/D$.

Beyond the domain of  applicability  of LRT, we have found a range of
values of the  parameters  for which the SNR exhibits a maximum for a
non-zero   noise  level.  Our  result  is  obtained  from   numerical
simulations by integrating  the  corresponding  Langevin  equation by
means of a standard  second-order  Runge-Kutta  method for stochastic
differential  equations
\cite{sde1,sde2}.
The SNR, computed  numerically  through the
averaged   power   spectrum,   is  defined   as
\begin{equation}
\label{eq:a5} {\rm SNR}=10\log{A(\omega_0)  \over N(\omega_0)} \;\; ,
\end{equation}
where $A(\omega_0)$ is the area of the peak above the noise floor and
$N(\omega_0)$  is the noise  background at the frequency  $\omega_0$.
In Fig.  \ref{fig2}  we have  depicted  the  ${\rm  SNR}$ for  finite
amplitudes of the oscillating field.  We have observed that for small
$\alpha$ the ${\rm SNR }$ is a monotonic  decreasing  function of the
noise, as  predicted  by LRT.  Nevertheless,  for  higher  values  of
$\alpha$  the  ${\rm  SNR  }$  has a  maximum  at $D  \neq  0$  (Fig.
\ref{fig2}).  When the amplitude decreases, this maximum becomes less
sharp and disappears at lower $\alpha$ in a continuous fashion.

This  behavior  could also be obtained  analytically  in the limit of
small  frequencies,  i.e.  when the system  relaxes  faster  than the
period  of the  external  field.  The  power  spectrum  may  then  be
computed  from  the  following   approximated   expression   for  the
correlation function
\begin{equation} \label{eq:a6}
\left<\cos(\theta_t)\cos(\theta_{t^\prime})\right> \approx
L_tL_{t^\prime}(1-e^{\lambda_{t,t^\prime}})
+\left(1-2{L_t\over \mu_t}\right)e^{\lambda_{t,t^\prime}} \;\;,
\end{equation}
where   $L_t   \equiv   L(\mu(t))$,    $\lambda_{t,t^\prime}   \equiv
-\int^{t^\prime}_t  \tau(\mu(s))^{-1}ds$  and $\mu_t \equiv  \mu(t)$.
Additionally, to obtain eq.  (\ref{eq:a6})  we have used the previous
analytical  expression of the relaxation time, which differs from the
exact  value only in a few per cent
\cite{Cof}.  The results  obtained  by using
eq.  (\ref{eq:a6})   in  eq.  (\ref{eq:a2})   are  depicted  in  Fig.
\ref{fig2}.  It is  interesting  to notice that the SNR defined  from
eq.  (\ref{eq:a2}),  which has dimensions of the inverse of time, has
been written in dimensionless units  
in such a way  that  it  coincides  with  the
definition of eq.  (\ref{eq:a5}).  The small differences  between the
analytical and simulation results occur at intermediate values of $D$
and come from the approximation in the relaxation time.

In the previous  results we have shown a dependence of the appearance
of  SR  on  the  amplitude  of  the  input  signal.  This  dependence
contrasts  with  the  results   obtained  for  the  bistable  quartic
potential,   in  which  the  maximum  of  the  SNR   disappears   for
sufficiently    large   amplitudes   of   the   oscillating    field.
Additionally,  numerical simulations show a qualitative change of the
SNR upon varying the frequency of the signal $\omega_0$.  The results
obtained  indicate  that  when  increasing  $\omega_0$,  for a  fixed
$\alpha$, the maximum of the SNR  disappears, as  illustrated in Fig.
\ref{fig3}.

Although,  in Fig.  \ref{fig3}  we have  depicted the ${\rm SNR}$ for
some  values of $\alpha > 1$, this  behavior  is similar  to that for
$\alpha < 1$, the reason  being the fact that when the dipole is able
to leave the  neighborhood of $\theta=0$ the noise is so high that it
destroys its coherent  motion.  However, fixed  $\omega_0$  for large
enough $\alpha$ the amount of noise necessary in order for the dipole
to leave the unstable equilibrium point does not destroy its coherent
motion.  This  behavior  is  illustrated  in  Fig.  \ref{fig4},   for
$\alpha=3$, $k=1$ and $\omega_0=0.1$
and is similar to the one found in \cite{aaay}.
It shows the time evolution of
$\cos(\theta)$  for the optimal, lower and greater noise levels.  For
the  previous  values of the  parameters,  the SNR (Fig.  \ref{fig5})
exhibits  a  maximum  at   $D=0.033$.  The  addition  of  noise  then
increases  the ${\rm  SNR}$ up to 15dB over its value at the limit of
$D$ going to zero.  For the optimal  value of $D$ the behavior of the
dipole looks surprisingly similar to that for the bistable potential.

Recalling our attention to Fig.  \ref{fig3}a, it also presents another
even more interesting  feature.  For low $D$ the ${\rm SNR}$ is not a
monotonic   increasing  function  of  the  amplitude.  This  kind  of
behavior,  presented  here for the first time,  implies  that in some
range of  values of the  parameters,  a small  signal is more  easily
detected than a greater one.  The output  signal,  however, is always
an  increasing  function  of the  amplitude.  The  maximum of the SNR
(Fig.  \ref{fig6}),  as a function  of  $\alpha$,  is due to the fact
that the noise at the frequency  $\omega_0$ increases faster than the
signal  when   $\alpha$   approaches   to  $1$.  For   $D=0.01$   and
$\omega_0=0.01$ the SNR has a maximum at $\alpha\approx0.7$.  In such
a case,  the  ${\rm  SNR}$  may be  evaluated  by  approximating  the
dynamics around the minimum by a  two-dimensional  Orsntein-Uhlenbeck
process \cite{diner}.
When this  approximation  holds  the  correlation  function
reads:
\begin{equation} \label{eq:a7}
\left<\cos(\theta_{t+\tau})\cos(\theta_t)\right> \approx
D^2{(1+e^{-\int_t^{t+\tau}h(s)ds})\over2h(t)h(t+\tau)} \;\; .
\end{equation}
The ${\rm SNR}$ has been  computed as we did with eq.  (\ref{eq:a6}).
The  corresponding  result  is  represented  in Fig.  \ref{fig6}.  In
spite of the  approximation  made to obtain eq.  (\ref{eq:a7}),  this
result  is in  excellent  agreement  with  the one  obtained  through
simulations for the range where this  approximation  makes sense.  It
must be pointed  out that the SNR, for low  amplitudes,  follows  the
quadratic  law, ${\rm SNR}  \propto  \alpha^2$.  Note that the result
given through eq.  (\ref{eq:a7})  corresponds to the one for a linear
system, which shows that  nonlinearities  do not play any role in the
appearance of this  phenomenon.  What is remarkable is the great deal
of generality of this result, since it may hold for any system around
a minimum.

In summary, we have shown the  existence  of SR in a dipole, a system
which has been  widely  analyzed  due to its  importance  as a simple
relaxation model.
It is well-known that the susceptibility of the dipole
exhibits a maximum at a certain value of the temperature
(strength of the noise) \cite{capi}.
In this paper we have also shown
that a maximum also appears in the SNR. 
The mechanism responsible for the appearance of SR
in this monostable system 
lies on the role  played  by noise  in the  relaxational  dynamics
itself  towards  the  stable  equilibrium   state.  As  an  important
feature, we have also found that the SNR depends on the  amplitude of
the  oscillating  field and  reaches a maximum at a certain  value of
this  parameter.  This result  reveals  the  existence  of an optimal
amplitude of the signal in order to be detected.  Although the dipole
is of interest in its own right, our results  could also be  extended
to the situation in which the system is around an  equilibrium  state
in a force field whose intensity  varies  periodically  in time.  

This work was  supported  by DGICYT of the Spanish  Government  under
Grant  No.  PB92-0859.  One  of  us  (JMGV)   wishes   to  thank  the
Generalitat de Catalunya for financial support.

\begin{figure}[th]
\caption[f]{\label{fig1}
(a) Potential $V(\theta)=-h(t)\cos(\theta)$ for the maximum of
h(t) (continuous line) and for the minimum
(dashed line), with $k=1$ and
$\alpha=0.5$. (b) Same plot, but with $\alpha=1.5$. }
\end{figure}

\begin{figure}[th]
\caption[e]{\label{fig2}
SNR as a function of the noise level $D$,
obtained through simulations (symbols) and computed from
eq. (\ref{eq:a6}) (lines). The values of the parameters
used here are $k=1$,
$\omega_0=0.01$, $\alpha=0.5$ (squares),
$\alpha=0.7$ (open circles) 
and $\alpha=0.9$ (filled circles).}
\end{figure}

\begin{figure}[th]
\caption[d]{\label{fig3}
Contour plot of the SNR as a function of both
the noise level $D$ and the amplitude of the
oscillating field $\alpha$,
for $k=1$ and the frequencies: (a) $\omega_0/2\pi=0.01$,
(b) $\omega_0/2\pi=0.1$ and (c) $\omega_0/2\pi=1$.}
\end{figure}

\begin{figure}[th]
\caption[c]{\label{fig4}
Time evolution of $\cos(\theta)$ for $\alpha=3$,
$k=1$, $\omega_0/2\pi=0.1$ and the noise levels:
$D=3.3\times10^{-2}$ (top),
$D=3.3\times10^{-6}$ (middle)
and $D=3.3$ (bottom).}
\end{figure}

\begin{figure}[th]
\caption[b]{\label{fig5}
SNR for $\alpha=3$,
$k=1$, $\omega_0/2\pi=0.1$.}
\end{figure}

\begin{figure}[th]
\caption[a]{\label{fig6}
SNR as a function of amplitude of the oscillating field
$\alpha$, for $k=1$, $\omega_0=0.01$ and $D=0.01$.
Obtained through simulations (symbols) and computed from
eq. (\ref{eq:a7}) (solid line).}
\end{figure}

\end{document}